\documentclass[useAMS,usenatbib]{mn2e}
\usepackage[colorlinks=true, linkcolor=blue, citecolor=blue, urlcolor=blue]{hyperref}
\usepackage{subfigure}
\usepackage{graphicx}
\usepackage{natbib}
\usepackage{psfig}

\def\aj{AJ}
\def\apj{ApJ}
\def\apjl{ApJL}
\def\apjs{ApJS}
\def\aap{A\&A}
\def\aaps{A\&AS}
\def\mnras{MNRAS}
\def\pasp{PASP}
\def\nat{Nature}

\catcode`\@=11
\def\gsim{\ifmmode{\mathrel{\mathpalette\@versim>}}
    \else{$\mathrel{\mathpalette\@versim>}$}\fi}
\def\lsim{\ifmmode{\mathrel{\mathpalette\@versim<}}
    \else{$\mathrel{\mathpalette\@versim<}$}\fi}
\def\@versim#1#2{\lower 2.9truept \vbox{\baselineskip 0pt \lineskip
    0.5truept \ialign{$\m@th#1\hfil##\hfil$\crcr#2\crcr\sim\crcr}}}
\catcode`\@=12

\newcommand{\beq}{\begin{equation}}
\newcommand{\eeq}{\end{equation}}

\begin{document}

\title[The tale of J1328$+$2752]
{Tale of J1328+2752: a misaligned double-double radio galaxy 
hosted by a binary black-hole?}
\author[Nandi et al.]
{S. Nandi$^{1}$$\thanks{e-mail: sumana@kth.se (SN)}$;
M. Jamrozy$^{2}$, R. Roy$^{3}$, J. Larsson$^{1}$, 
\newauthor
D.J. Saikia$^{4,5}$, M. Baes$^{6}$ and  M. Singh$^{7}$\\
 $^1$KTH, Department of Physics, and the Oskar Klein Centre, AlbaNova, SE-106 91 Stockholm, Sweden\\
 $^2$Obserwatorium Astronomiczne, Uniwersytet Jagiello\'nski, ul. Orla 171, 
30-244 Krak\'ow, Poland\\
 $^3$The Oskar Klein Centre, Department of Astronomy, Stockholm University, AlbaNova, 10691 Stockholm, Sweden\\
  $^4$Cotton College State University, Panbazar, Guwahati 781 001, India\\
  $^5$National Centre for Radio Astrophysics, TIFR, Pune University Campus, Post Bag 3, Pune 411 007, India\\
  $^6$Sterrenkundig Observatorium, Universiteit Gent, Krijgslaan 281 S9, B-9000 Gent, Belgium\\
  $^7$Aryabhatta Research Institute of Observational Sciences (ARIES), Manora Peak, Nainital, 263 129, India\\
 }

\date{Accepted ??; Received ???}

\pagerange{\pageref{firstpage}--\pageref{lastpage}} \pubyear{}

\maketitle

\label{firstpage}

\begin{abstract}
 We present a radio and optical study of the double-double radio 
 galaxy J1328+2752 based on new low-frequency GMRT observations 
 and SDSS data. The radio data were used to investigate
 the morphology and to perform a spectral index analysis. 
 In this source we find that the inner double is misaligned by 
 $\sim$30$^\circ$ from the axis of the outer diffuse structure.
 The SDSS spectrum shows that the central component has double-peaked 
 line profiles with different emission strengths. The average velocity 
 off-set of the two components is 235$\pm$10.5~km~s$^{-1}$. 
 The misaligned radio morphology along with the double-peaked 
 emission lines indicate that this source is a potential candidate 
 binary supermassive black hole. This study further supports mergers 
 as a possible explanation for repeated jet activity in radio sources.     
\end{abstract}


\section{Introduction}
\label{section1}

 The existence of two or more pairs of synchrotron emitting radio lobes
 driven by the same central active galactic nucleus (AGN) is extremely 
 important for understanding the evolution of AGNs, as it provides vital 
 evidence for multiple episodes of nuclear activity. Such sources are often 
 termed double-double radio galaxies (DDRGs) \citep{2000MNRAS.315..371S}. 
 These DDRGs exhibit a wide range of linear sizes, from less than few 
 hundreds of kpc up to more than a Mpc 
 \citep{2009BASI...37...63S, 2012BASI...40..121N, 2013MNRAS.436.1595K}.
 For these episodic sources the new jets usually follow the same 
 direction as the previous jets. There are a few examples of 
 ``misaligned DDRGs'' which have different orientation of axes 
 for the two epochs. \citet{2006MNRAS.366.1391S} show that the 
 misalignment angle is within $\sim$20$^\circ$ for a sample of 
 12 DDRGs. However, the steep-spectrum core-dominated radio galaxy 
 3C293 provides excellent observational evidence of restarted jet 
 activity along with a $\sim$35$^\circ$ jet rotation 
 \citep{1996MNRAS.278....1A}. The estimated time-scale  of episodic 
 activity in 3C293 is $\sim$10$^5$~yr, which is significantly smaller 
 than most other known DDRGs \citep{2011MNRAS.414.1397J}. 
 In accordance with this, \citet{2016A&A...595A..46M} found that the outer 
 lobes are $\sim$60 Myr old and that the jet activity related to the formation 
 of the outer lobes ceased within the last Myr. Meanwhile, the misaligned inner 
 lobes are only about $\sim$0.3 Myr old. The derived jet time-scales provide
 strong support for the Lense-Thirring precession model 
 \citep{1975ApJ...195L..65B}, in which the supermassive black hole (SMBH) spin,
 and therefore the jet axis, flips rapidly. Recently, 
 \citet{2013MNRAS.436..690S} identified another DDRG, B0707-359, with a 
 misalignment angle of about $\sim$30$^\circ$ between the two epochs 
 of activity.
 
\begin{figure*}
\vbox{

\hbox{
      \psfig{file=325MHZ_J1328_1.PS,width=2.2in,angle=0}%
      \psfig{file=610MHZ_J1328_1.PS,width=2.25in,angle=0}
       \psfig{file=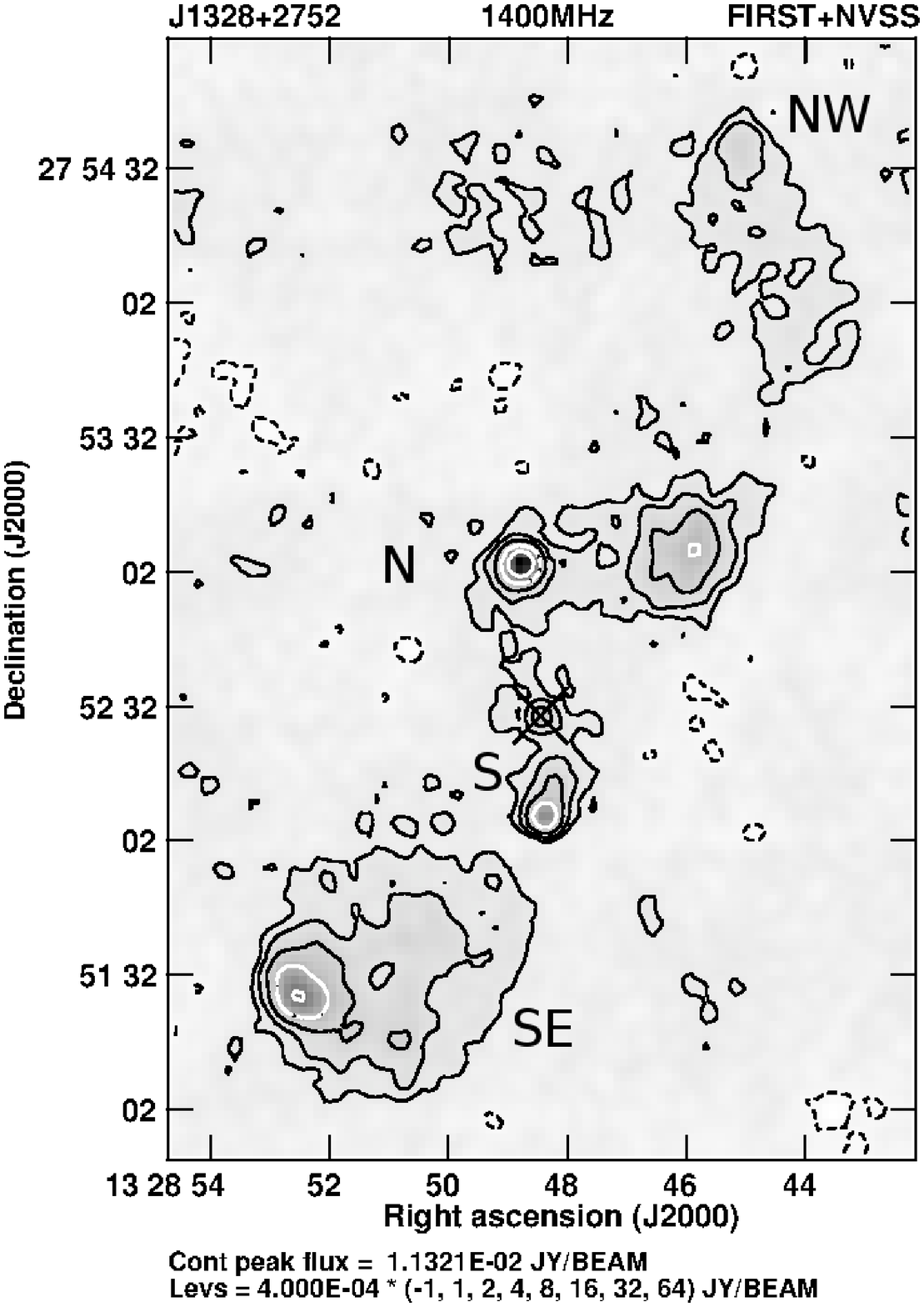,width=2.0in,angle=0}     
      }
}
\caption[]{Images of J1328+2752 at 325, 607 and 1400 MHz.
The position of the optical host is marked with a `$\times$' sign. The 
locations of the different components are marked in the 1400 MHz map. The image details are presented in Table \ref{table2}.}
\label{J1328_fig1}
\end{figure*}
\begin{figure}
\vbox{
\center 
\psfig{file=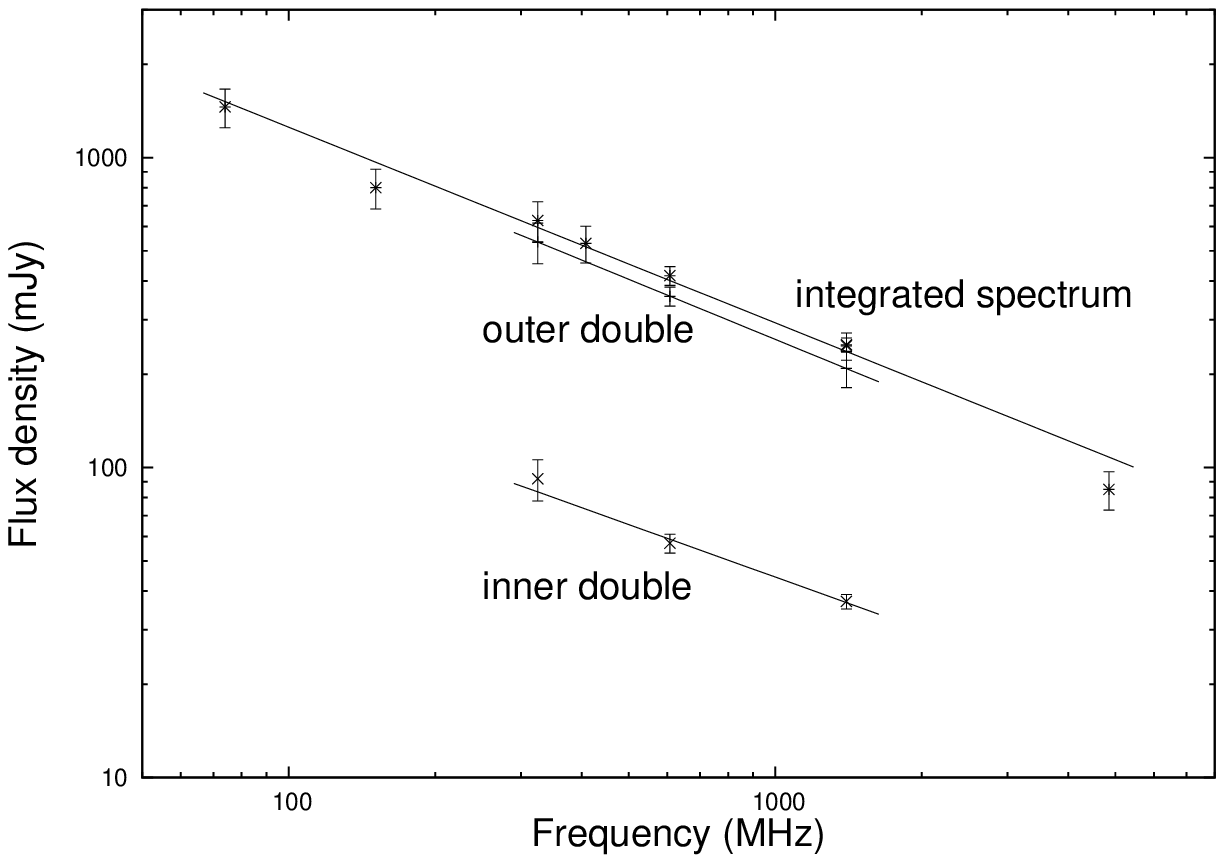,width=3.1in,angle= 0}
 \psfig{file=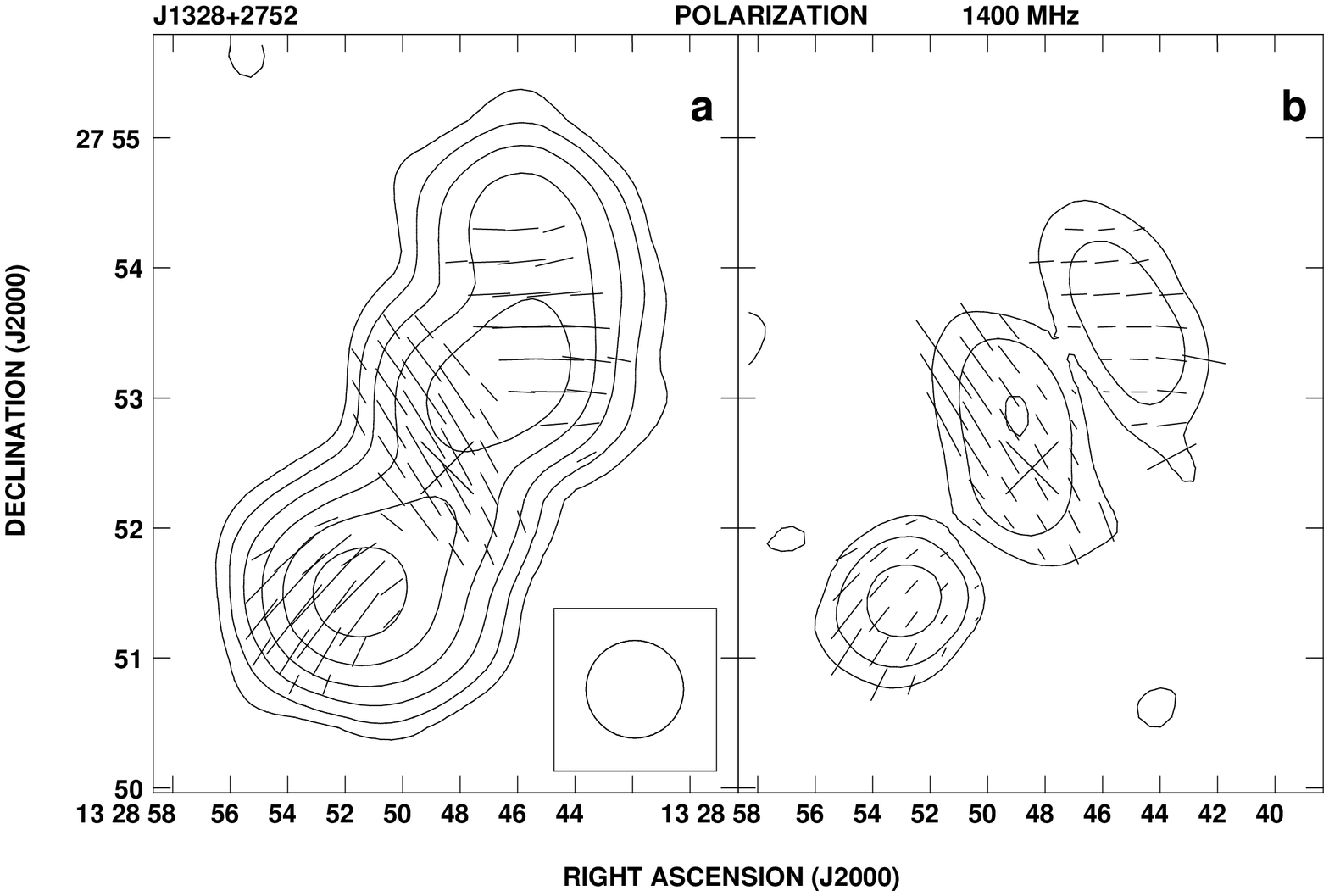,width=3.1in,angle= 0}
 \psfig{file=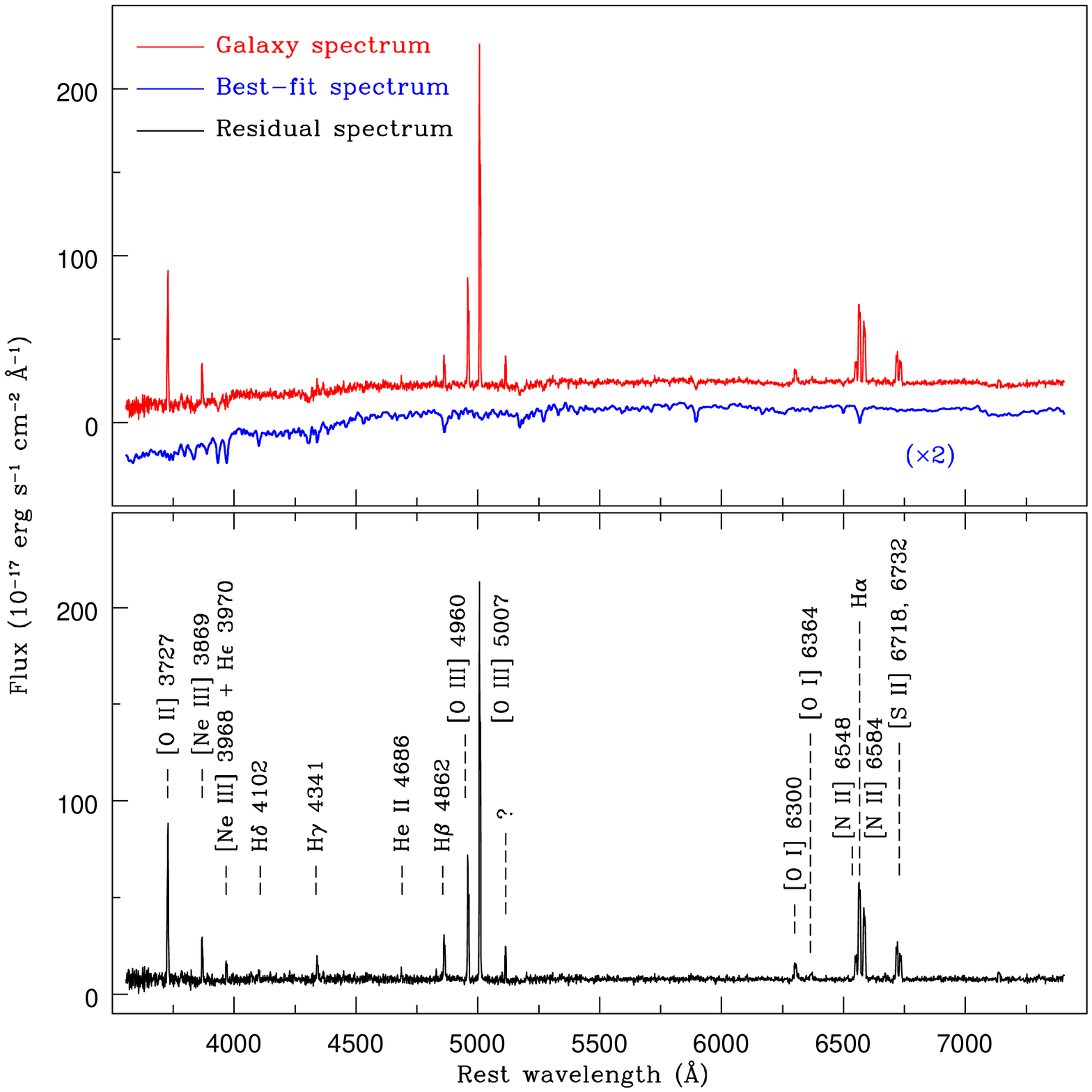,width=3.1in,angle= 0}
     }
\caption[]{Upper panel: Radio spectra of the whole source and the individual 
components. Middle panel: 1400 MHz NVSS polarimetric images. (a) Total 
intensity contours spaced by a factor of 2 starting at 1.35 mJy/beam. 
Superimposed are E-vectors with their lengths proportional to the polarized 
intensity, where 10 arcsec correspond to 1 mJy/beam. (b) Linearly polarized 
intensity contours spaced by a factor of 2 starting with 0.9 mJy/beam, with 
the vectors of the fractional linear polarization superimposed. A length of 
10 arcsec corresponds to 10 per cent of the fractional linear polarization. 
The `$\times$' sign marks the position of the host galaxy.
Lower panel: Redshift-corrected SDSS spectrum of the galaxy 
(in red) and the best-fit model spectrum for the underlying 
stellar population of the host produced by $pPXF$ (in blue)(upper half).
Lower panel: The residual spectrum with the strongest emission lines labelled 
(lower half).}
\label{J1328_fig2}
\end{figure}

 The reorientation of the jet axis in these misaligned DDRGs may be caused by 
 an axis precession or axis flip of the AGN. Influence of a nearby galaxy or 
 the coalescence of massive black holes may trigger a new jet with sufficient 
 axis rotation to explain the observations \citep{2015ApJ...810L...6R}. 
 Historically, radio galaxies with a rapid change of jet direction,  
 in particular ``X-shaped'' galaxies, have been proposed as candidates for 
 binary black holes (see e.g. \citealt{1980Natur.287..307B, 2001A&A...380..102M, 2007MNRAS.374.1085L, 2015ApJ...810L...6R}). These sources are also promising 
 contributers to the gravitational wave background 
 \citep{1980Natur.287..307B, 2016PhRvL.116x1103A}. Recently, double-peaked 
 emission lines in AGN have been suggested as an indicator for binary black 
 holes. Such lines are a possible signature of a bound pair of SMBHs, moving 
 with their own characteristic velocities 
 \citep{2011ApJ...733..103F, 2014MNRAS.437...32W, 2015ApJ...799..161K}.
 However, apart from binary black holes, there are other scenarios that may 
 explain double-peaked emission lines, including jet-cloud interactions or 
 rotating gaseous disks \citep{2012ApJ...752...63S}. 
 Of these different possibilities, the scenario of binary black holes is more 
 likely in the case of merger remnants or elliptical galaxies 
 \citep{2014Natur.511...57D}. A misaligned DDRG hosted by a giant elliptical 
 with double-peaked emission lines is therefore of great interest  
 as a potential candidate for harbouring a binary black hole. Such systems 
 may provide direct observational evidence for galaxy mergers as triggers 
 of multiple epochs of jet activity \citep{2014ApJ...789...16N}. 

  From our previous study \citep{2012BASI...40..121N} we have identified 
 one such source, J1328$+$2752. This system shows not only restarted jet 
 activity with an axis reorientation, but also double-peaked emissions lines 
 from the central AGN at a redshift 0.0911. \citet{2012ApJS..201...31G}
 included the host galaxy of J1328$+$2752 in a list of 3030 galaxies that show 
 double-peaked narrow emission-lines in their `Sloan Digital Sky Survey' 
 (SDSS) spectra. The linear sizes of the inner and outer doubles of this 
 source are 96 and 413 kpc, respectively. In this paper we present new
 low-frequency radio observations of J1328$+$2752 along with a study of
 the optical spectrum. We assume a Universe with 
 H$_0$=71 km s$^{-1}$ Mpc$^{-1}$, $\Omega_{\rm m}$=0.27 and $\Omega_{\rm vac}$=0.73.

\section{Observations and data analysis}
 Both radio and optical data have been used to analyse the characteristics
 of J1328$+$2752. 

 {\bf Radio data:} The radio observations were performed with the Giant 
 Metrewave Radio Telescope (GMRT) under proposal codes 23$\_$056 and 28$\_$044.  The target was observed on 2013-03-24 for 3.6 hr at 607 MHz and on 2015-06-27 
 for 4.7 hr at  325 MHz. We observed the flux density calibrator, 3C286, at 
 the beginning and end of each observing run and also used it as a bandpass 
 calibrator based on the scale of \citet{1977A&A....61...99B}. The phase 
 calibrator, J1330+251, was observed for $\sim$5 min after each of several 
 $\sim$20~min exposures of J1328$+$2752. We used the AIPS software package 
 for data reduction. To obtain the best possible images, several rounds of 
 self-calibration were performed. 
 
 For this study we also used images at 1400 MHz obtained from  FIRST (Faint 
 Images of the Radio Sky at Twenty centimeters; \citealt{1995ApJ...450..559B}) 
 and NVSS (NRAO VLA Sky Survey; \citealt{1998AJ....115.1693C}).
 Since about 40\% of the flux density (and structure) is lost in the 
 high-resolution FIRST map, the FIRST and NVSS images were combined to 
 image the range of structure shown in Fig. \ref{J1328_fig1}.
 The final merged map was checked for flux density consistency by comparing the
 flux-density of point sources in the FIRST and merged maps. This difference 
 did not exceed 1\%. The integrated flux densities at different 
 frequencies are given in Table \ref{table1}. Additionally, we used the NVSS 
 Q and U Stokes maps and the AIPS task COMB to create images of the linearly 
 polarized intensity and fractional polarization.

 {\bf Optical data:} The optical spectrum of J1328$+$2752 was obtained 
 from data release DR12 of  the SDSS. The observed spectrum is contaminated
 by the stellar absorption features of the host galaxy and also affected by the
 foreground extinction and recession velocity of the system. Before analyzing 
 the narrow emission lines, we have therefore removed all other features using 
 the penalized pixel-fitting tool ($pPXF$, \citealt{2004PASP..116..138C}). 
 
 \section{Results}
 The full-resolution maps at 325, 607 and 1400 MHz are shown in Fig. 
 \ref{J1328_fig1}. All these images show a new, inner structure centered at the
 optical host but about $\sim$30$^\circ$ off the axis of the earlier diffuse 
 emission. The central radio core is visible at 1400 MHz. For the GMRT images
 weak emission from the core is detected at 607 MHz, while the 325 MHz map does
 not show any clear core emission. The flux density values (see 
 Table \ref{table2}) of the central component indicate that this is a 
 flat-spectrum radio core. The outer lobes are asymmetric in morphology.
 A bright hotspot in the southern outer lobe is visible at all three 
 frequencies, while there is no evidence of a hotspot in the northern outer
 lobe. In the GMRT images we note a sharp bend of the northern outer lobe 
 towards the north-east and an enhancement of flux density at the position 
 where the lobe starts to bend. The combined 1400 MHz image also shows an 
 enhancement of flux density at the same position. The asymmetric inner double 
 is completely embedded in the older diffuse emission. The northern inner 
 component is dominated by a bright hotspot, while the southern inner 
 component has an elongated structure with a bright hotspot at the end. 
 The integrated flux densities and those of the components are presented in 
 Tables \ref{table1} and \ref{table2} respectively and plotted in 
 Fig. \ref{J1328_fig2}, upper panel. The estimated spectral index 
 $\alpha$ ($S_\nu\propto\nu^{-\alpha}$) for the whole source using 
 only the low-resolution values at 1400 MHz is 0.63$\pm$0.03.
 Ensuring that no flux density is missing, the spectral indices of the outer
 and inner doubles between 325 and 1400 MHz are 0.64$\pm$0.002 and 
 0.56$\pm$0.06 respectively. 
 
 The total intensity NVSS map with the electric field E-vectors superimposed 
 is shown in Fig. \ref{J1328_fig2}, middle panel (a). The Fig. 
 \ref{J1328_fig2}, middle panel (b) shows the linearly polarized intensity map 
 with the vectors of fractional linear polarization superimposed. It can be 
 seen that the whole structure is polarized. There are three distinct regions 
 visible in the polarimetric maps. The total integrated polarized flux 
 density of this source is $17.6\pm1.4$ mJy, which gives $\sim 7$ \%
 for the mean fractional polarization. 

\begin{figure*}
\vbox{
\hbox{
      \psfig{file=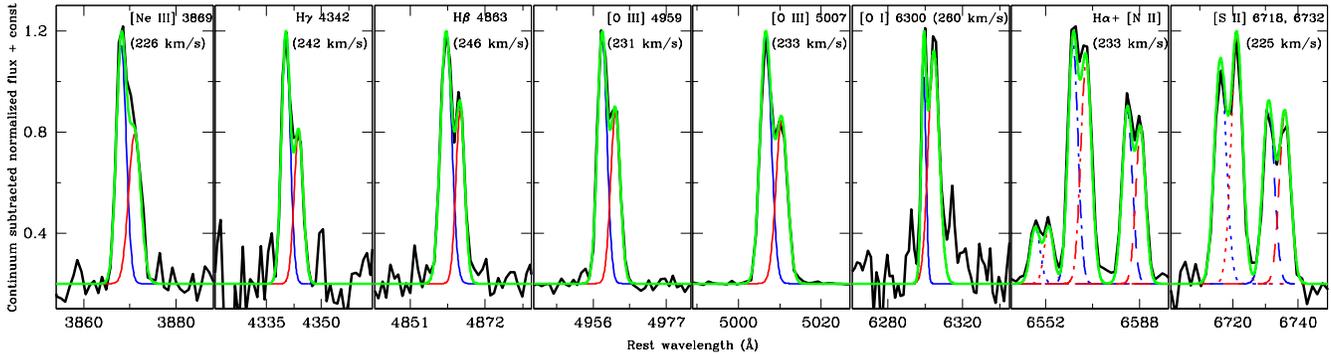,width=7.0in,angle=0}%

      }
}
\caption[]{The double-peaked emission lines obtained from the
de-reddened SDSS spectrum of J1328+2752. The velocity separations 
determined for each pair of lines are included in the upper,
right corners of the panels.}
\label{doublpeak}
\end{figure*}

\begin{figure}
\hbox{
 \psfig{file=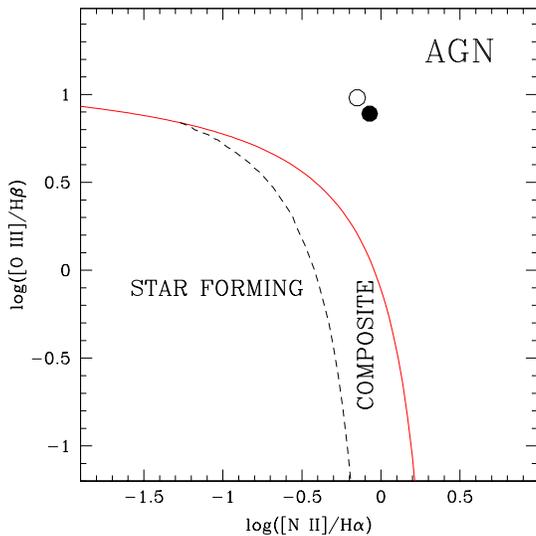,width=2.8in,angle=0}
     }
\caption[]{BPT diagram of J1328+2752. Open and filled 
circles represent the blue and red components of the system. 
The solid line separates star forming galaxies and AGNs. Composite galaxies 
occupy the region between the dashed and solid lines.}
\label{BPT}
\end{figure}

 The optical spectrum (Fig. \ref{J1328_fig2}, lower panel) shows that 
 J1328+2752 has an elliptical host with high-ionization narrow emission 
 lines. The typical stellar velocity dispersion of the host determined 
 using pPXF is 193.1$\pm$13.6 kms$^{-1}$. This is consistent with the result 
 from the SDSS DR12 survey (velocity dispersion 176.8$\pm$9.7 kms$^{-1}$). 
 A close inspection shows that the strong emission lines 
 of J1328+2752 are double peaked. Each continuum-subtracted emission line was
 deblended with two Gaussian functions using the IRAF routine
 SPLOT\footnote{IRAF is distributed by the National Optical Astronomy 
 Observatory, which is operated by the Association of Universities for 
 Research in Astronomy, Inc., under cooperative agreement with the National 
 Science Foundation.}. The continuum-subtracted normalized flux of
 eleven major emission lines are shown together with the normalized model 
 in Fig. \ref{doublpeak}. 
 The lines are  [Ne III] $\lambda3869$, H$\gamma$ $\lambda4342$, H$\beta$
 $\lambda4863$, [O III] $\lambda4959$, [O III] $\lambda5007$, [O I] 
 $\lambda6300$, H$\alpha$ $\lambda6563$, [N II] $\lambda\lambda6548,6584$, 
 and [S II] $\lambda\lambda6718,6732$. The left (blue) and right (red) 
 Gaussian may correspond to two merging components. The total model
 flux (green) of each spectral line is also shown in the Fig. \ref{doublpeak}. 
 After fitting these lines we find that the two components are separated by a 
 velocity of 235 $\pm$ 10.5 kms$^{-1}$. Such a velocity separation between 
 the blue and red components is consistent with a scenario of two cores in 
 merging galaxies \citep{2014MNRAS.437...32W}. For every line, the intensities 
 of the two narrow-line components are different. To understand 
 the classes of the two merging components, we use the BPT diagnostic diagram 
 \citep{1981PASP...93....5B, 2006MNRAS.372..961K} shown in Fig. \ref{BPT}. This 
 analysis distinguishes AGN charateristics from those of starforming and 
 composite galaxies using line ratios. 
 The [N II]/H$\alpha$ and [O III]/H$\beta$ ratios calculated for the two 
 components place them at nearby locations in the BPT diagram. This makes the 
 possibility of `jet-cloud interaction' quite unlikely and almost rules out
 the `rotating-disk' scenario \citep{2012ApJ...752...63S}. It also indicates 
 that this is a system of two merging AGNs. 
 
 The stellar velocity dispersion can be used to estimate the black hole mass 
 through the relation given by \citet{2002ApJ...574..740T}. For J1328$+$2752, 
 the SDSS velocity dispersion gives a mass for the central black hole (or black
 hole pair) of $\log(\rm M_{BH}/M_{\odot}) = 7.91\pm0.17$.
 This value is not as large as most SMBH masses in powerful radio galaxies,
 although objects of similar masses do exist \citep{2011MNRAS.415.1013K}.

\begin{table}
   \caption{Integrated flux densities}
   \begin{tabular}{ccccc}
  \hline
  Obs. Freq.  &S             &error          &survey  & Ref\\
  (MHz)       &(mJy)         &(mJy)        & used   & used\\
  (1)         &(2)            &(3)           &(4)     & (5) \\
 \hline
  74          &1457           &208           &VLSSr    &(1)\\
  151         &800            &117           &7C      &(2)\\
  325         &627            &94            &GMRT    &(3)\\
  408         &529            &72            &B2      &(4)\\
  607         &414            &29            &GMRT    &(3)\\ 
 1400         &247            &25            &GB      &(5)\\
 1400         &249            &13            &NVSS    &(6)\\
 1400         &152            &8             &FIRST   &(7)\\   
 1400         &206            &10            &        &(8)\\
4850         &85             &12             &87GB    &(9)\\ 
\hline
\end{tabular}
\\
Column 1 gives the frequency, Columns 2 and 3 give the total flux densities of the source and the flux density error, Column 4 gives the name of the survey, Column 5 gives references- (1) \citet{2014MNRAS.440..327L} 
(The original flux density of the VLSSr survey was multiplied by a factor of 0.9 to be consistent 
with the \citet{1977A&A....61...99B} scale), (2) \citet{1996MNRAS.282..779W}, (3) Our observation,
(4) \citet{1972A&AS....7....1C}, (5) \cite{1992ApJS...79..331W}, (6) \citet{1998AJ....115.1693C},
(7) \citet{1997ApJ...475..479W}, (8) combined NVSS and FIRST from present paper,
 (9) \cite{1991ApJS...75.1011G}. 
\label{table1}
\end{table}

\begin{table}
\caption{Observational parameters and flux densities.}
\begin{tabular}{l rrr c l r r}
\hline
Freq.        & \multicolumn{3}{c}{Beam size}   & rms &  Cmp. & S$_p$  & S$_t$    \\
(MHz)        & ($^{\prime\prime}$) & ($^{\prime\prime}$) &($^\circ$) &   (mJy &  & (mJy   &  (mJy)  \\
             &         &       &       & /b)  &        &  /b)    \\ 
(1)          & (2)     & (3)   & (4)    & (5) &  (6)   & (7) & (8)     \\ 
\hline
  G325       & 13.15   & 7.18  & 69  &0.38 & NW  & 19.5  & 239     \\
             &         &       &   &     & N   & 37    &  44    \\
             &         &       &   &     & C   &       & $\lsim$1.5    \\
             &         &       &   &     & S   & 21    &  48     \\
             &         &       &   &     & SE  & 39    &  296    \\
                          
  G607       & 5.27    & 4.22  & 69  &0.07 & NW  & 3.5   & 164       \\
             &         &       &   &     & N   & 20    &  28       \\
             &         &       &   &     & C   & 1.8      &2      \\
             &         &       &   &     & S   & 11    &  29    \\
             &         &       &   &     & SE  & 11    &  193    \\
 
  V1400      & 5.40    & 5.40  & 0  &0.19 & NW  & 3.3     &73      \\
             &         &       &   &     & N   & 11     &19       \\
             &         &       &   &     & C   & 2      &2       \\
             &         &       &   &     & S   & 5      &18            \\
             &         &       &   &     & SE  & 7     &95          \\    

\hline
\end{tabular}
Column 1: frequency of observations where G and V indicate GMRT and VLA 
respectively; Columns 2 to 4: beam size in arcsec and its PA in degrees; 
Column 5: rms in units of mJy/beam; Column 6: component designation, where 
NW, N, S and SE denote north-west outer, north inner, south innner and 
south-east outer components, respectively. C represents the central component. 
Columns 7, 8: the peak flux densities and total flux densities measured 
from the total intensity maps for each component in units of mJy/beam and mJy, 
respectively.\\ 
\label{table2}
\end{table}

\section{DISCUSSIONS AND CONCLUSIONS}
 We have presented 325 and 607 MHz GMRT continuum observations
 of J1328+2752. These data, together with the 1400 MHz image, reveal that the 
 source has four distinct components in addition to a flat-spectrum core.   
 The spectral indices of the outer double is steeper than the inner one, which 
 indicate two epochs of jet activity 
 (e.g. \citealt{2012MNRAS.424.1061K, 2013MNRAS.436.1595K,
 2015A&A...584A.112O}) although the difference here is small due to an active 
 hot-spot in the southern outer lobe and possible re-acceleration in the sharp 
 bend in the northern outer lobe. From the radio morphology and spectral index 
 analysis we confirm this source as a DDRG with an axis rotation. The 
 percentage of polarized emission is similar to that observed for X-shaped 
 radio galaxies which show jet reorientation
 \citep[e.g.,][]{2012MNRAS.422.1546K}. 
 Here we speculate that the different strengths of the two components of 
 the optical emission lines is an indication that the system is associated with
 two distinct narrow line regions (NLRs) orbiting around each of two separate
 nuclei, plausibly constituted by a merging massive black hole binary.
 This source may therefore be a relevant candidate for future gravitational 
 wave experiments. Moreover, this study supports the scenario that renewed jet 
 activity may be associated with merger events. However, this dual black hole 
 system cannot be spatially resolved by current available data. We plan for
 Very Long Baseline Interferometry (VLBI) imaging and high angular resolution
 optical imaging to reveal the existence of a massive dual/binary black hole.
 It is possible to identify radio-emitting binary black holes with parsec-scale
 separations through VLBI \citep{2014Natur.511...57D}. A linear separation by 
 more than 8 or 3 pc of presumably two radio cores of J1328+2752 can be 
 revealed, respectively, through 1600 or 5000 MHz VLBI observations.

\section*{Acknowledgments}
We would like to thank an anonymous referee for many useful 
comments. SN is funded by Wenner-Gren foundation (Stockholm,
Sweden) to conduct her research projects. 
SN is thankful to Belgian Federal Science Policy (BELSPO) for financial 
assistance. M.J. acknowledges support by the Polish NSC grant No. 
2013/09/B/ST9/00599. We thank the GMRT staff for technical support during 
the observations. GMRT is run by the National Centre for Radio Astrophysics 
of the Tata Institute of Fundamental Research. 
Funding for the SDSS and SDSS-II has been provided by
the Alfred P. Sloan Foundation, the Participating Institutions,
the National Science Foundation, the U.S. Department of
Energy, the National Aeronautics and Space Administration,
the Japanese Monbukagakusho, the Max Planck Society, and
the Higher Education Funding Council for England.



\begin{thebibliography}{39}
\expandafter\ifx\csname natexlab\endcsname\relax\def\natexlab#1{#1}\fi

\bibitem[{{Abbott} {et~al}\mbox{.}(2016){Abbott}, {Abbott}, {Abbott},
  {Abernathy}, {Acernese}, {Ackley}, {Adams}, {Adams}, {Addesso}, {Adhikari},
  \& et~al.}]{2016PhRvL.116x1103A}
{Abbott} B.~P. {et~al.}, 2016, Physical Review Letters, 116, 241103

\bibitem[{{Akujor} {et~al}\mbox{.}(1996){Akujor}, {Leahy}, {Garrington},
  {Sanghera}, {Spencer}, \& {Schilizzi}}]{1996MNRAS.278....1A}
{Akujor} C.~E., {Leahy} J.~P., {Garrington} S.~T., {Sanghera} H., {Spencer}
  R.~E., {Schilizzi} R.~T., 1996, \mnras, 278, 1

\bibitem[{{Baars} {et~al}\mbox{.}(1977){Baars}, {Genzel}, {Pauliny-Toth}, \&
  {Witzel}}]{1977A&A....61...99B}
{Baars} J.~W.~M., {Genzel} R., {Pauliny-Toth} I.~I.~K., {Witzel} A., 1977,
  \aap, 61, 99

\bibitem[{{Baldwin} {et~al}\mbox{.}(1981){Baldwin}, {Phillips}, \&
  {Terlevich}}]{1981PASP...93....5B}
{Baldwin} J.~A., {Phillips} M.~M., {Terlevich} R., 1981, \pasp, 93, 5

\bibitem[{{Bardeen} \& {Petterson}(1975)}]{1975ApJ...195L..65B}
{Bardeen} J.~M., {Petterson} J.~A., 1975, \apjl, 195, L65

\bibitem[{{Becker} {et~al}\mbox{.}(1995){Becker}, {White}, \&
  {Helfand}}]{1995ApJ...450..559B}
{Becker} R.~H., {White} R.~L., {Helfand} D.~J., 1995, \apj, 450, 559

\bibitem[{{Begelman} {et~al}\mbox{.}(1980){Begelman}, {Blandford}, \&
  {Rees}}]{1980Natur.287..307B}
{Begelman} M.~C., {Blandford} R.~D., {Rees} M.~J., 1980, \nat, 287, 307

\bibitem[{{Cappellari} \& {Emsellem}(2004)}]{2004PASP..116..138C}
{Cappellari} M., {Emsellem} E., 2004, \pasp, 116, 138

\bibitem[{{Colla} {et~al}\mbox{.}(1972){Colla}, {Fanti}, {Fanti}, {Ficarra},
  {Formiggini}, {Gandolfi}, {Lari}, {Marano}, {Padrielli}, \&
  {Tomasi}}]{1972A&AS....7....1C}
{Colla} G. {et~al.}, 1972, \aaps, 7, 1

\bibitem[{{Condon} {et~al}\mbox{.}(1998){Condon}, {Cotton}, {Greisen}, {Yin},
  {Perley}, {Taylor}, \& {Broderick}}]{1998AJ....115.1693C}
{Condon} J.~J., {Cotton} W.~D., {Greisen} E.~W., {Yin} Q.~F., {Perley} R.~A.,
  {Taylor} G.~B., {Broderick} J.~J., 1998, \aj, 115, 1693

\bibitem[{{Deane} {et~al}\mbox{.}(2014){Deane}, {Paragi}, {Jarvis}, {Coriat},
  {Bernardi}, {Fender}, {Frey}, {Heywood}, {Kl{\"o}ckner}, {Grainge}, \&
  {Rumsey}}]{2014Natur.511...57D}
{Deane} R.~P. {et~al.}, 2014, \nat, 511, 57

\bibitem[{{Fu} {et~al}\mbox{.}(2011){Fu}, {Myers}, {Djorgovski}, \&
  {Yan}}]{2011ApJ...733..103F}
{Fu} H., {Myers} A.~D., {Djorgovski} S.~G., {Yan} L., 2011, \apj, 733, 103

\bibitem[{{Ge} {et~al}\mbox{.}(2012){Ge}, {Hu}, {Wang}, {Bai}, \&
  {Zhang}}]{2012ApJS..201...31G}
{Ge} J.-Q., {Hu} C., {Wang} J.-M., {Bai} J.-M., {Zhang} S., 2012, \apjs, 201,
  31

\bibitem[{{Gregory} \& {Condon}(1991)}]{1991ApJS...75.1011G}
{Gregory} P.~C., {Condon} J.~J., 1991, \apjs, 75, 1011

\bibitem[{{Joshi} {et~al}\mbox{.}(2011){Joshi}, {Nandi}, {Saikia},
  {Ishwara-Chandra}, \& {Konar}}]{2011MNRAS.414.1397J}
{Joshi} S.~A., {Nandi} S., {Saikia} D.~J., {Ishwara-Chandra} C.~H., {Konar} C.,
  2011, \mnras, 414, 1397

\bibitem[{{Kewley} {et~al}\mbox{.}(2006){Kewley}, {Groves}, {Kauffmann}, \&
  {Heckman}}]{2006MNRAS.372..961K}
{Kewley} L.~J., {Groves} B., {Kauffmann} G., {Heckman} T., 2006, \mnras, 372,
  961

\bibitem[{{Kharb} {et~al}\mbox{.}(2015){Kharb}, {Das}, {Paragi}, {Subramanian},
  \& {Chitta}}]{2015ApJ...799..161K}
{Kharb} P., {Das} M., {Paragi} Z., {Subramanian} S., {Chitta} L.~P., 2015,
  \apj, 799, 161

\bibitem[{{Konar} \& {Hardcastle}(2013)}]{2013MNRAS.436.1595K}
{Konar} C., {Hardcastle} M.~J., 2013, \mnras, 436, 1595

\bibitem[{{Konar} {et~al}\mbox{.}(2012){Konar}, {Hardcastle}, {Jamrozy},
  {Croston}, \& {Nandi}}]{2012MNRAS.424.1061K}
{Konar} C., {Hardcastle} M.~J., {Jamrozy} M., {Croston} J.~H., {Nandi} S.,
  2012, \mnras, 424, 1061

\bibitem[{{Kozie{\l}-Wierzbowska} {et~al}\mbox{.}(2012){Kozie{\l}-Wierzbowska},
  {Jamrozy}, {Zola}, {Stachowski}, \& {Ku{\'z}micz}}]{2012MNRAS.422.1546K}
{Kozie{\l}-Wierzbowska} D., {Jamrozy} M., {Zola} S., {Stachowski} G.,
  {Ku{\'z}micz} A., 2012, \mnras, 422, 1546

\bibitem[{{Kozie{\l}-Wierzbowska} \&
  {Stasi{\'n}ska}(2011)}]{2011MNRAS.415.1013K}
{Kozie{\l}-Wierzbowska} D., {Stasi{\'n}ska} G., 2011, \mnras, 415, 1013

\bibitem[{{Lal} \& {Rao}(2007)}]{2007MNRAS.374.1085L}
{Lal} D.~V., {Rao} A.~P., 2007, \mnras, 374, 1085

\bibitem[{{Lane} {et~al}\mbox{.}(2014){Lane}, {Cotton}, {van Velzen}, {Clarke},
  {Kassim}, {Helmboldt}, {Lazio}, \& {Cohen}}]{2014MNRAS.440..327L}
{Lane} W.~M., {Cotton} W.~D., {van Velzen} S., {Clarke} T.~E., {Kassim} N.~E.,
  {Helmboldt} J.~F., {Lazio} T.~J.~W., {Cohen} A.~S., 2014, \mnras, 440, 327

\bibitem[{{Machalski} {et~al}\mbox{.}(2016){Machalski}, {Jamrozy}, {Stawarz},
  \& {We{\.z}gowiec}}]{2016A&A...595A..46M}
{Machalski} J., {Jamrozy} M., {Stawarz} {\L}., {We{\.z}gowiec} M., 2016, \aap,
  595, A46

\bibitem[{{Murgia} {et~al}\mbox{.}(2001){Murgia}, {Parma}, {de Ruiter},
  {Bondi}, {Ekers}, {Fanti}, \& {Fomalont}}]{2001A&A...380..102M}
{Murgia} M., {Parma} P., {de Ruiter} H.~R., {Bondi} M., {Ekers} R.~D., {Fanti}
  R., {Fomalont} E.~B., 2001, \aap, 380, 102

\bibitem[{{Nandi} {et~al}\mbox{.}(2014){Nandi}, {Roy}, {Saikia}, {Singh},
  {Chandola}, {Baes}, {Joshi}, {Gentile}, \& {Patgiri}}]{2014ApJ...789...16N}
{Nandi} S. {et~al.}, 2014, \apj, 789, 16

\bibitem[{{Nandi} \& {Saikia}(2012)}]{2012BASI...40..121N}
{Nandi} S., {Saikia} D.~J., 2012, Bulletin of the Astronomical Society of
  India, 40, 121

\bibitem[{{Orr{\`u}} {et~al}\mbox{.}(2015){Orr{\`u}}, {van Velzen}, {Pizzo},
  {Yatawatta}, {Paladino}, {Iacobelli}, {Murgia}, {Falcke}, {Morganti}, {de
  Bruyn}, {Ferrari}, {Anderson}, {Bonafede}, {Mulcahy}, {Asgekar}, {Avruch},
  {Beck}, {Bell}, {van Bemmel}, {Bentum}, {Bernardi}, {Best}, {Breitling},
  {Broderick}, {Br{\"u}ggen}, {Butcher}, {Ciardi}, {Conway}, {Corstanje}, {de
  Geus}, {Deller}, {Duscha}, {Eisl{\"o}ffel}, {Engels}, {Frieswijk}, {Garrett},
  {Grie{\ss}meier}, {Gunst}, {Hamaker}, {Heald}, {Hoeft}, {van der Horst},
  {Intema}, {Juette}, {Kohler}, {Kondratiev}, {Kuniyoshi}, {Kuper}, {Loose},
  {Maat}, {Mann}, {Markoff}, {McFadden}, {McKay-Bukowski}, {Miley}, {Moldon},
  {Molenaar}, {Munk}, {Nelles}, {Paas}, {Pandey-Pommier}, {Pandey}, {Pietka},
  {Polatidis}, {Reich}, {R{\"o}ttgering}, {Rowlinson}, {Scaife},
  {Schoenmakers}, {Schwarz}, {Serylak}, {Shulevski}, {Smirnov}, {Steinmetz},
  {Stewart}, {Swinbank}, {Tagger}, {Tasse}, {Thoudam}, {Toribio}, {Vermeulen},
  {Vocks}, {van Weeren}, {Wijers}, {Wise}, \& {Wucknitz}}]{2015A&A...584A.112O}
{Orr{\`u}} E. {et~al.}, 2015, \aap, 584, A112

\bibitem[{{Roberts} {et~al}\mbox{.}(2015){Roberts}, {Saripalli}, \&
  {Subrahmanyan}}]{2015ApJ...810L...6R}
{Roberts} D.~H., {Saripalli} L., {Subrahmanyan} R., 2015, \apjl, 810, L6

\bibitem[{{Saikia} \& {Jamrozy}(2009)}]{2009BASI...37...63S}
{Saikia} D.~J., {Jamrozy} M., 2009, Bulletin of the Astronomical Society of
  India, 37, 63

\bibitem[{{Saikia} {et~al}\mbox{.}(2006){Saikia}, {Konar}, \&
  {Kulkarni}}]{2006MNRAS.366.1391S}
{Saikia} D.~J., {Konar} C., {Kulkarni} V.~K., 2006, \mnras, 366, 1391

\bibitem[{{Saripalli} {et~al}\mbox{.}(2013){Saripalli}, {Malarecki},
  {Subrahmanyan}, {Jones}, \& {Staveley-Smith}}]{2013MNRAS.436..690S}
{Saripalli} L., {Malarecki} J.~M., {Subrahmanyan} R., {Jones} D.~H.,
  {Staveley-Smith} L., 2013, \mnras, 436, 690

\bibitem[{{Schoenmakers} {et~al}\mbox{.}(2000){Schoenmakers}, {de Bruyn},
  {R{\"o}ttgering}, {van der Laan}, \& {Kaiser}}]{2000MNRAS.315..371S}
{Schoenmakers} A.~P., {de Bruyn} A.~G., {R{\"o}ttgering} H.~J.~A., {van der
  Laan} H., {Kaiser} C.~R., 2000, \mnras, 315, 371

\bibitem[{{Smith} {et~al}\mbox{.}(2012){Smith}, {Shields}, {Salviander},
  {Stevens}, \& {Rosario}}]{2012ApJ...752...63S}
{Smith} K.~L., {Shields} G.~A., {Salviander} S., {Stevens} A.~C., {Rosario}
  D.~J., 2012, \apj, 752, 63

\bibitem[{{Tremaine} {et~al}\mbox{.}(2002){Tremaine}, {Gebhardt}, {Bender},
  {Bower}, {Dressler}, {Faber}, {Filippenko}, {Green}, {Grillmair}, {Ho},
  {Kormendy}, {Lauer}, {Magorrian}, {Pinkney}, \&
  {Richstone}}]{2002ApJ...574..740T}
{Tremaine} S. {et~al.}, 2002, \apj, 574, 740

\bibitem[{{Waldram} {et~al}\mbox{.}(1996){Waldram}, {Yates}, {Riley}, \&
  {Warner}}]{1996MNRAS.282..779W}
{Waldram} E.~M., {Yates} J.~A., {Riley} J.~M., {Warner} P.~J., 1996, \mnras,
  282, 779

\bibitem[{{White} \& {Becker}(1992)}]{1992ApJS...79..331W}
{White} R.~L., {Becker} R.~H., 1992, \apjs, 79, 331

\bibitem[{{White} {et~al}\mbox{.}(1997){White}, {Becker}, {Helfand}, \&
  {Gregg}}]{1997ApJ...475..479W}
{White} R.~L., {Becker} R.~H., {Helfand} D.~J., {Gregg} M.~D., 1997, \apj, 475,
  479

\bibitem[{{Woo} {et~al}\mbox{.}(2014){Woo}, {Cho}, {Husemann}, {Komossa},
  {Park}, \& {Bennert}}]{2014MNRAS.437...32W}
{Woo} J.-H., {Cho} H., {Husemann} B., {Komossa} S., {Park} D., {Bennert} V.~N.,
  2014, \mnras, 437, 32

\end{thebibliography}
\end{document}